# Direct Observation of Whispering Gallery Mode Polariton and Its Dispersion in a ZnO tapered Microcavity

Liaoxin Sun[1], Zhanghai Chen[1,*], Qijun Ren[1], Ke Yu[2], Lihui Bai[1], Weihang Zhou[1], Hui Xiong[1], Z. Q. Zhu[2], Xuechu Shen[1,†]

1. Surface Physics Laboratory, Department of Physics, Fudan University, Shanghai 200433, China

2. Department of Electronic Engineering, East China Normal University, Shanghai 200062, China

Abstract

We report direct observation of the strong exciton-photon coupling in ZnO tapered whispering gallery (WG) microcavity at room temperature. By scanning excitations along the tapered arm of ZnO tetrapod using micro-photoluminescence spectrometer with different polarizations, we observed a transition from the pure WG optical modes in the weak interaction regime to the excitonic polariton in the strong coupling regime. The experimental observations are well described by using the plane wave model including excitonic polariton dispersion relation. This provides a direct mapping of the polariton dispersion, and thus a comprehensive picture for coupling of different excitons with differently polarized WG modes.

[*] zhanghai@fudan.edu.cn

[†] xcshen@fudan.edu.cn

Semiconductor microcavities have fascinated many researchers in the last decade because they allow precise control of the light-matter interaction.[1,2,3] Planar microcavities consisting of quantum wells surrounded by Bragg reflectors have been fabricated to investigate the coupling between excitons and optical modes. A strong photon-exciton coupling leads to the formation of a new kind of quasi-particle, *i.e. exciton polariton,* which has an anticrossing behavior in its dispersion diagram[4]. The use of Fabry-Pérot (FP) microcavities, mostly GaAs based structures partially due to its nearly perfect crystal quality, has already brought us a significant physical understanding of excitonic polariton physics[5,6]. Recent development of polariton physics has two major and fundamentally important directions, one is the realization of Bose-Einstein condensation for polariton[7], and the other is the development of polariton based novel devices such as "polariton laser"[8,9]. They require a large coupling strength and/or room temperature operation, which unfortunately can not be fulfilled by the GaAs based planar cavities. This is because these materials do not exhibit strong exciton oscillator strength, and the exciton binding energy is relatively small. Moreover, the overlap of exciton with cavity modes is rather weak due to the narrowness of active quantum well layer. On the other hand, one probably can avoid these problems by fabricating ZnO based microcavities, because ZnO is a wide band gap semiconductor having large exciton oscillator strength and binding energy (~60 meV). Besides, ZnO is itself a promising material for UV or visible photonic nanodevices[10,11]. Indeed, a large Rabi splitting of 120 meV has been reported recently for exciton polaritons in ZnO nanowire cavity exhibiting a FP type resonance[12]. This finding strongly suggests that ZnO microcavities may be a promising system for studying polariton physics and for developing "polariton laser". However fabrication of high quality planar resonator using ZnO has not been achieved thus far[13].

Currently, the research of ZnO microcavity has been focused on the so-called whispering gallery (WG) resonator[18]. Taking advantage of hexagonal cross section perpendicular to the longitudinal crystal axis (c-axis) of ZnO, fabrication of hexagonally shaped WG resonator is easily achievable by using chemical methods. WG modes (WGMs) in a single ZnO nanoneedle have been investigated in the defect related visible emission band, where polariton effect is negligible and WGMs remain pure optical wave behavior in this energy regime[14]. The resonant interaction between exciton and WG modes, i.e. the polariton effect in a WG resonator, which is of great importance for developing polariton laser, has not yet been examined. In this Letter, we report for the first time, a strong photon-exciton coupling, i.e. the formation of excitonic polariton, in a ZnO WG resonator. Direct mapping of polariton dispersions is achieved by using the spatially resolved spectroscopic technique. Calculation based on the plane wave model and polariton dispersion relation fits well with the experimental observations. Our results provide a comprehensive picture for coupling of excitons with differently polarized WG modes. We show that ZnO WG resonator has superior qualities for investigating excitonic polaritons and developing novel polariton based devices.

The tapered WG resonator used is one of the arms of the tetrapod ZnO structures synthesized on silicon substrate by a vapor-phase transport method at temperature of ~700 °C under the atmospheric pressure[15]. Figure 1 shows scanning electron micrographs (SEMs) of a ZnO tetrapod. It can be seen that the tetrapod has arm lengths of ~30 μm with a diameter continuously decreasing from several micrometers at the root to a few nanometers at the tip top. SEM image in Fig. 1(b) shows that the tetrapod used here has flat and smooth hexagonal resonator facets, and thus the optical loss due to roughness of the cavity faces was suppressed. Since the tapered arm is actually a varying radius WG resonator in which the energy of WG modes and the orders of resonance can be modulated, the energy difference between the cavity modes and excitons can be tuned and controlled by choosing the radius of the resonator. WG resonator with lateral size comparable to the wavelength of the excitonic light-emission, can generate largely separated modes in the UV range. These allow us to investigate the interaction of single WG mode of different orders with excitons. In addition, unlike the planar cavities, the overlap between exciton and cavity mode is greatly enhanced in a WG resonator, because the body of a WG resonator is itself an active medium for excitonic emission. Indeed, Kaliteevski suggested[16] that the overlap of cavity mode and exciton can be close to unity for WG resonators. By performing a scanning excitation along the tapered arm of ZnO tetrapod using micro-photoluminescence spectrometer with polarization dependence, one can expect a transition from the pure WG optical modes in the weak interaction regime to the excitonic polariton in the nearly resonant regime.

The spectroscopic experiments are carried out in a confocal micro-photoluminescence system, using a pulsed laser line of 371 nm as the excitation source. The laser spot focused on the sample is sub-µm and the scanning step was set to 0.2 µm along the ZnO nanorod axle which is parallel to its **c**-axis. We performed emission spectral measurements for detections of unpolarized, TM polarized (the electrical component of light ***E***//**c**-axis), and TE polarized (***E***⊥**c**-axis) signals, respectively, while the electrical vector of excitation laser E is always parallel to **c**-axis.

Typical spectra of the ZnO tetrapod in the UV emission range of interband transitions taken at room temperature are presented in Fig. 2. The black curve is the spectrum measured at the center of the ZnO tetrapod, where the cavity has not been formed. This spectrum shows a broadened peak at 3.220 eV with a long low energy tail. As proposed before, the origin of this UV emission is mainly due to exciton transitions and their phonon replicas[10]. It is this UV emission band that allows us to systematically observe the interaction of cavity modes with exciton directly by use of the micro-photoluminescence experiment while scanning the excitation spot along the tapered rod. The conduction band of wurtzite ZnO is a s-like state with $\Gamma_7$ symmetry, whereas the valence band is p-like state which is split into three bands due to effects of crystal-field and spin-orbit interactions. Reynolds *et al.* assigned the three bands ordering as A-$\Gamma_9$, B-$\Gamma_7$, and C-$\Gamma_7$ through polarized optical reflectance and magnetoluminescence experiments[17]. According to the selection rules for one-photon process of optical transitions, all three excitonic transitions are allowed in the σ polarization ($E \perp$ **c**-axis and $k \perp$ **c**-axis), although the C excitonic emission is quite weak. The C excitonic transition is allowed in the π polarization with $E$//**c**-axis and $k \perp$ **c**-axis, while the transition is forbidden for A exciton, and only weakly observed for B exciton in this geometry[18]. Excitons with different symmetries are expected to interact with differently polarized cavity modes. As shown in Fig. 2(a), these are confirmed in the spectra measured at a certain position of the tapered arm of ZnO, where the WG resonator is possible, with different polarization configurations. The three peaks marked with dashed lines result from the WG resonance. To confirm this, and more importantly, to observe the polariton luminescence, we performed the micro-luminescence scan in unpolarized and polarized detection. The results are shown in Fig. 2(b) and Fig. 3, respectively. Due to the aforementioned selection rules, these results give the luminescence of different polariton modes.

It can be seen from Fig. 2(b) which is for the measurements with unpolarized detection, that the UV band of luminescence is clearly modulated during the scanning. It shows a blueshift of the spectral maxima with the gradual decrease of the diameter of the tapered arm. The shift is not linear and appears anticrossing behavior at the energy range of 3.23 eV~3.30 eV. Moreover, the peak blueshift is periodic, that is, second and third period appear when further scanning along the tapered rod. We plot in Fig. 2(b) three periods of the spectra for which the laser spot scans near the middle part of the arm. The periodic behavior of the WGMs was also reported in Ref. [14] for defect related visible light emission, but the blueshift of the modes bends to opposite direction showing no anticrossing and the pure optical plane wave behavior. Here, we attribute the observation of strong off-linear behavior of WGMs to the anticrossing effect of the polariton dispersion. Since the energy of the UV interband transition luminescence is closed to the photon energies of the WGM modes in the microcavity, therefore the observed luminescence is dominated by excitonic polaritons resulting from a strong coupling between WGMs and excitons in the cavity, though it may be more WGM-like in the low energy side while more exciton-like in the high energy side of the energy period ranged from ~3.10 to ~3.30 eV. The bending and strong off-linearity at the photon energy of 3.235 eV in the spectra curves indicate a transition of the hybridized mode behavior from WGM-like to exciton-polariton like. As the laser spot scans along the **c**-axis from the bottom to the tip of the ZnO resonator, the WG modes with different interference orders enter UV region one by one to couple with the excitons, giving rise to the observed periodic behavior of the polariton luminescence. Obviously, at least two sets of WG modes with different periods can be resolved in Fig. 2(b). These two sets of WG modes are well defined when we detect the emission with TE or TM polarization, as shown in Fig. 3. According to the selection rules, the behavior of the WG mode shown in Fig. 3(a) (and the corresponding waterfall plot Fig. 3(c)) represents the evolution of A and B excitonic polariton as a function of cavity size. The result given in Fig. 3(b) (and the corresponding waterfall plot Fig. 3(d)) is quite different in its spread range and bending position, and reveals actually the evolution of C excitonic polariton during the scanning. We notice that the C excitonic transition is allowed while A and B excitonic transitions are forbidden or only weakly allowed in TM configuration, and its transition (recombination) energy is around 46 meV and 40 meV higher than those of A and B excitons, repectively[12], which makes the strong bending of its blue shift curve occur at higher energy than that for A and B.

By scanning the excitation spot along the **c**-axis in different polarization configuration, the dispersions of A, B and C excitonic polariton are mapped, respectively. Although the diameter of resonator is comparable with the wavelength of UV excitonic emission of ZnO, we can still use a plane wave model to describe the experimental results. This model has been successfully applied to describe the WGMs within the visible spectral range in a ZnO needle [14]. The radius of circumscribing circle of the resonator $R$ can be written as:

$$R = \frac{hc}{3\sqrt{3}nE}[N + \frac{6}{\pi}\arctan(\beta\sqrt{3n^2 - 4})] \quad . \tag{1}$$

Here, $n$ is the refractive index, $h$ the Planck constant and c the speed of light. The integer $N$ is the interference order of the resonance. The factor β reflects polarization, $\beta=n$ for TE mode and $\beta=1/n$ for TM mode. In the UV interband transition region, cavity modes are strongly coupled to free excitons resulting in the excitonic polariton. This effect modulates dramatically the dielectric function as well as the refractive index $n$ of the system. This situation is very different from the visible emission region as reported in Ref. 14. It is well known that the polariton dispersion relation in the strong coupling regime can be expressed as[12]:

$$n^2 = \varepsilon(\omega,k) = \varepsilon_\infty(1 + \sum_{j=A,B,C} \Omega_j \frac{\omega_{j,L}^2 - \omega_{j,T}^2}{\omega_{j,T}^2 - \omega^2 - i\omega\gamma_j}) = \frac{c^2k^2}{\omega^2} \tag{2}$$

with $\varepsilon_\infty$ being a background dielectric constant, $\omega_{j,T}$ and $\omega_{j,L}$ being the transverse and longitudinal resonance frequencies at $k=0$, and $\gamma_j$ being the damping constants. The prefactor $\Omega_j$ is defined by J. Lagois[19]. By taking the WGM resonance condition and the polariton dispersion into account, one can get an exact expression for the resonance energy as a function of radius $R$ in this hexagonal resonator via refractive index $n$. We carried out a computational analysis by using Eqs. (1) and (2), and give the calculated results (solid curves) in Fig. 3.

As mentioned above, A- and B- excitonic polaritons are detected in the TE polarization detection, while the C-excitonic polariton is detected in TM geometry. The parameters chosen in the calculation are summarized in Table I (can also be found in Ref. [8, 12, 19]). The damping term is neglected and Rabi splitting is independent with R in our calculation [16]. As illustrated by Fig. 3, the theoretical calculations for the lower polariton branches give an accurate fit to the experiments, which leads to a conclusion that the polaritons observed in the hexagonal cavity of the tapered arm of ZnO were the coupled and hybridized modes of excitons A, B and C with the WGMs with mode number N=6, 7, 8, respectively. It is remarkable that the WGMs can be observed up to such high mode numbers and their couplings with different excitons are clearly resolved even in the room temperature micro-PL measurement. For clarity, we present here complete theoretical dispersions with and without polariton coupling for TE and TM, respectively [see Figs. 3(e) and figs. 3(f)]. As mentioned above, the observed emission is from the lower polariton branch and the upper polariton is absent. One reasonable explanation of the absence of the upper polariton branch is that the interaction between polariton and phonon via the exciton component is very strong at room temperature, and the energy of the upper branch is hundreds of meV higher than that of the lower branch; therefore, the polaritons at the upper branch are quickly decayed to the lower energy band; hence, the emission form upper polariton branch is strongly suppressed.

To further understand the WGM polariton behavior and thus highlight its possible impact on optoelectronic application, we explore physical mechanism for the intensity evolution of the polariton emission. As described by Hopfield[20], the probability of the polariton escape from the crystal as photon emission is directly proportional to its group velocity $v_p$. From $v_p = (1/\hbar)\partial E/\partial k$ and the polariton dispersion function $E=E(k)$, one can conclude that the $v_p$ would be greatly reduced as the hybridized mode approaches the free exciton energy position. In this case, it becomes difficult for the polariton to propagate to the surface of the crystal and then escape as photon emission. In Fig. 2(b) and Fig. 3, when the mode behavior is transited from more WGM-like to more exciton-like along with the mode frequency entering into the energy regime of resonant photon-exciton interaction, the intensity of the polariton emission begins to decrease and eventually disappears (the peaks disappeared above 3.27 eV for the TE and above 3.31 eV for the TM polarization, respectively), exactly as predicted by the above picture of polariton group velocity. These results indicate that one may manipulate the polariton group velocity by tuning the WGM-exciton coupling strength via tailoring the resonator dimension. This maybe useful for developing novel polariton based devices.

In summary, we have reported a direct observation of the interaction between WGM and free exciton with polarization dependence. We demonstrated that the TE polarization and the TM polarization cavity modes couple with different dipole orientated free excitons, i.e. presented the comprehensive picture for the coupling of A, B, C excitons with differently polarized WG modes. The dispersion relation of the hybridized modes is mapped out. The results indicate that WG resonator of ZnO may be an ideal system for the investigation of polariton physics and the developing of novel polariton based devices.

The work is funded by the NSF of China, 973 projects of China (No. 2004CB619004 and No. 2006CB921506) . The authors are grateful to Professors H. Guo, D. L. Feng, C. M. Hu for stimulating discussions.

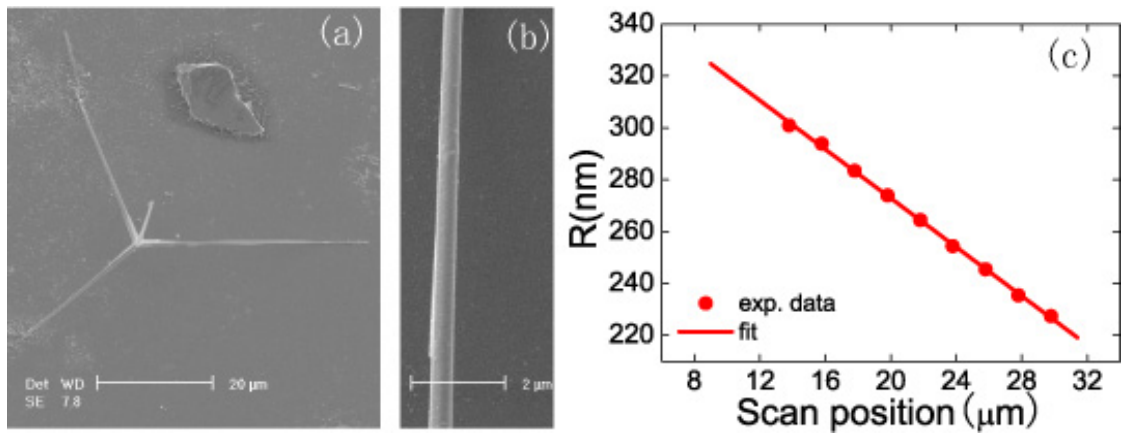

**FIG. 1.** （a） **SEM image of the ZnO tetrapod. (b) Magnified SEM image of the resonator shows perfect surface quality. (c) The relation of the measured radius *R* and scan position *x* is fitted by** $R = 367 - 4.7 \times x$ **(nm).**

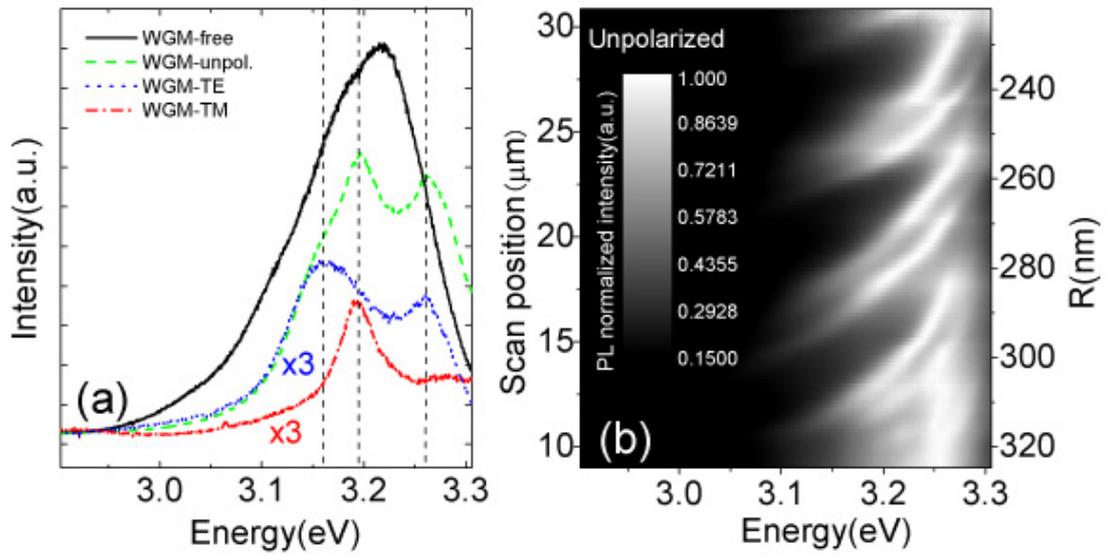

**FIG. 2. (a) WGM-free (black curve) spectrum and the WGM spectra obtained at a certain position of the tapered arm with unpolarized (green dashed curve), TE (blue dotted curve) and TM polarized (red dash-dotted curve) detection, respectively. (b) Spatial-resolved PL mapping along the tapered arm with the excitation laser polarization E//c-axis and detection unpolarized.**

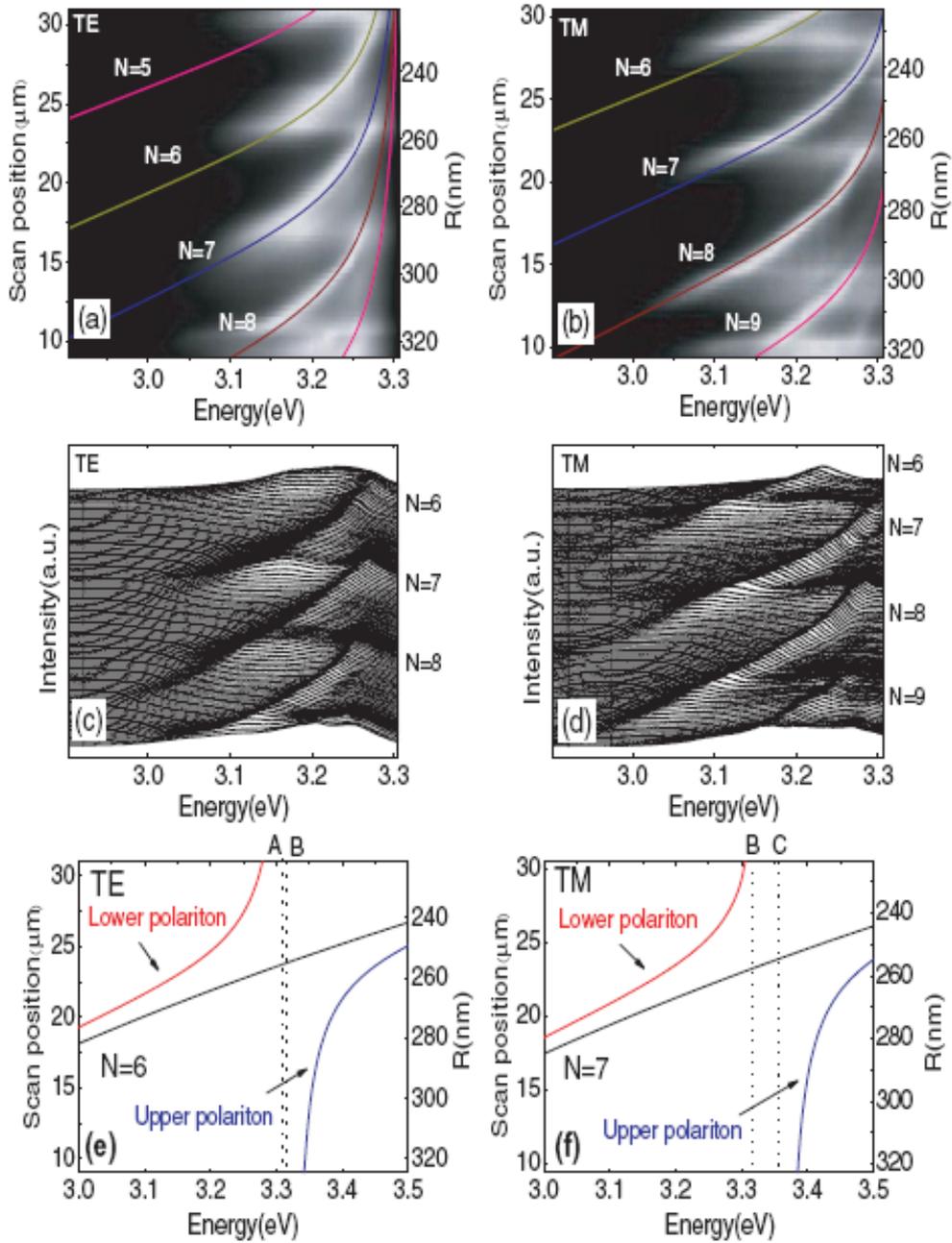

**FIG. 3.** PL mapping along the ZnO tapered arm with the excitation polarization E//c-axis and detection polarizations of (a) TE and (b) TM. (c) and (d) are the corresponding waterfall plots of the spectra. The colored lines in (a) and (b) are the theoretical fittings. The PL intensity scale of the mappings (a) and (b) is the same as that of Fig. 2(b). (e) and (f) the complete theoretical

**computation dispersion for TE and TM, respectively. The energy of A-, B- and C-exciton were shown with dotted lines, the lower (red), upper polariton branch (blue), pure WGM (black) are shown.**

TABLE I. Parameters of excitons chosen in the calculation, here, $\omega_{TL} = \omega_L - \omega_T$

|  | A exciton | B exciton | C exciton |
|---|---|---|---|
| $\hbar\omega_T$(eV) | 3.309 | 3.315 | 3.355 |
| $\hbar\omega_{TL}$(meV) | 2.0 | $10.4^{TE}$, $0.8^{TM}$ | 12.0 |
| $\varepsilon_\infty$ |  | 6.2 |  |